\documentstyle[aps,prl,epsf,floats]{revtex} 

\newcommand{\nn}{\nonumber}

\def\Dsl{\hbox{/\kern-.6000em D}} %roman D

\def\dsl{\,\raise.15ex\hbox{/}\mkern-13.5mu D}
\def\bsigma{\mbox{\boldmath $\sigma$}}

\def\bsigma{\mbox{\boldmath $\sigma$}}

\def\ltap{\ \raise.3ex\hbox{$<$\kern-.75em\lower1ex\hbox{$\sim$}}\ }
\def\gtap{\ \raise.3ex\hbox{$>$\kern-.75em\lower1ex\hbox{$\sim$}}\ }
\def\OMIT#1{}

\begin{document}

\twocolumn[\hsize\textwidth\columnwidth\hsize\csname@twocolumnfalse\endcsname

\preprint{\tighten  \hbox{UCSD/PTH 00-06}
%       \hbox{hep-ph/00xxx}
}

\title{Logarithms of $\alpha$ in QED bound states from the renormalization
group}

\author{Aneesh V. Manohar\footnote{amanohar@ucsd.edu} and
Iain W.\ Stewart\footnote{iain@schwinger.ucsd.edu} \\[4pt]}
\address{\tighten Department of Physics, University of California at San
Diego,\\[2pt] 9500 Gilman Drive, La Jolla, CA 92099 }

\maketitle

{\tighten
\begin{abstract}

The velocity renormalization group is used to determine $\ln \alpha$
contributions to QED bound state energies. The leading order anomalous
dimension for the potential gives the $\alpha^5 \ln \alpha$ Bethe logarithm in
the Lamb shift. The next-to-leading order anomalous dimension determines the
$\alpha^6 \ln \alpha$, $\alpha^7 \ln^2\! \alpha$, and $\alpha^8 \ln^3\! \alpha$
corrections to the energy. These are used to obtain the $\alpha^8 \ln^3\! \alpha$
Lamb shift and $\alpha^7 \ln^2\! \alpha$ hyperfine splitting for Hydrogen,
muonium and positronium, as well as the $\alpha^2 \ln \alpha$ and $\alpha^3
\ln^2\! \alpha$ corrections to the ortho- and para-positronium lifetimes.

\end{abstract}
}%end tighten
]\narrowtext %% end two-column

The energies of QED bound states such as positronium, muonium or Hydrogen
depend on $\ln \alpha$~\cite{pachucki}. For example, the Lamb shift\cite{Lamb}
of Hydrogen contains the famous Bethe logarithm of order $\alpha^5 \ln \alpha$.
In relativistic scattering at high energy $E\gg m$, logarithms of $E$ can be
determined from renormalization group equations. The leading logarithmic series
of the form $(\alpha \ln m/E )^n$ can be summed by integrating the one-loop
anomalous dimension, the subleading series $\alpha (\alpha \ln m/E )^n$ by
integrating the two-loop anomalous dimension, and so on. In this letter, we
show how to predict logarithms of $\alpha$ for non-relativistic QED bound
states using the velocity renormalization group (VRG)~\cite{LMR}. Our approach
can also be used for other non-relativistic systems.

The leading order (LO) anomalous dimension generates the Bethe logarithm of
order $\alpha^5 \ln \alpha$ for the Lamb shift, and is the only term in the
series. Integrating the next-to-leading order (NLO) anomalous dimension
determines the $\alpha^6 \ln\alpha$, $\alpha^7 \ln^2\! \alpha$ and $\alpha^8
\ln^3\! \alpha$ terms, and simultaneously gives the Lamb shift and hyperfine
splitting for Hydrogen, muonium and positronium, as well as the decay widths
for ortho- and para-positronium. Here we derive all these terms except the
$\alpha^7 \ln^2\! \alpha$ Lamb shift and $\alpha^6 \ln \alpha$ Lamb shift and
hyperfine splitting, which depend on terms in the NLO anomalous dimension for
which values are not presented here. The $\alpha^8\ln^3\alpha$ term for
Hydrogen has been the subject of a recent debate in the literature. An analytic
calculation by Karshenboim~\cite{karshenboim} and a numerical calculation by
Goidenko et al.~\cite{goidenko} agree with each other, but disagree with
numerical calculations by Mallampalli and Sapirstein~\cite{ms} and by
Yerokhin~\cite{yerokhin} (which agree). The renormalization group answer agrees
with the analytic $\alpha^8\ln^3\! \alpha$ result of Karshenboim, and we will
comment on the disagreement with Refs.~\cite{ms,yerokhin}. The 
$\alpha^8\ln^3\! \alpha$ positronium Lamb shift is a new result.

Our calculation makes use of the non-relativistic effective theory for
QED~\cite{caswell} (NRQED), formulated as in Refs.~\cite{LMR,amis}. We will
consider the interaction of two particles of mass $m_1$ and $m_2$, and charge
$-e$ and $Ze$ respectively. The effective theory has a subtraction velocity
$\nu$, rather than the usual subtraction scale $\mu$ of dimensional
regularization. QED is matched onto the effective theory at $\nu=1$, and the
effective Lagrangian is scaled down to $\nu=Z \alpha$, the typical velocity in
the bound state, using the VRG. (Terms $\propto \ln(m_2/m_1)$ are suppressed
for $m_2\gg m_1$ and are neglected.) Finally, the energy is computed from
matrix elements of the Lagrangian.  In this approach, all logarithms arise from
renormalization group running, since matrix elements of operators renormalized
at $\nu=Z \alpha$ and matching coefficients at $\nu=1$ contain no large
logarithms. This should be compared with the traditional approach where $\ln
\alpha$ terms are determined by examining integrals for the matrix elements.
Determining the logarithms using the VRG gives them a universal and simple
description.

The interaction potential has an expansion in powers of the velocity $v$,
\vskip-0.1truecm
\begin{eqnarray}
  V = V^{(-1)} + V^{(0)} + V^{(1)} + V^{(2)}+ V^{(3)} + \ldots \,,
\end{eqnarray}
\vskip-0.1truecm\noindent
where $V^{(n)}\sim v^n$. The lowest order interaction is the Coulomb potential,
$V^{(-1)}$, of order $\alpha/v$. [In momentum space $V^{(-1)}\sim 1/{\bf
k}^2\sim v^{-2}$. Shifting the power by one makes the power counting simpler.]
The potentials in the center of mass frame for scattering of particles with
momenta $\pm \bf p$ to $\pm \bf p^\prime$, with $\bf k = p^\prime -p$ are
\vskip-0.1truecm
\begin{eqnarray} \label{Vpp}
&&  V^{(-1)} = {{U}_c \over {\mathbf k}^2} \,, \qquad
 V^{(0)} ={ {U}_k  \over |{\mathbf k}| } \,, \\
&& V^{(1)} = U_2 + U_s\: {\bf S^2} + { U_r ({\mathbf p^2 + p^{\prime 2}}) \over
  2 {\mathbf k}^2} - {i {\mathbf U}_\Lambda \cdot ({\mathbf p^\prime \times p})
  \over  {\mathbf k}^2 } \nn\\
&& \qquad\quad  + U_t \Big( {\mathbf \bsigma_1 \cdot \bsigma_2}-{3\,{\mathbf k
 \cdot \bsigma_1}\,  {\mathbf k \cdot \bsigma_2} \over {\mathbf k}^2} \Big )
  \nn \,,  \\
&& V^{(2)}\!=\! \left(U_{3}\!+\!U_{3s}{\bf S}^2\right) |{\mathbf k}|
   \! + \!\ldots , \
  V^{(3)}\!=\! {U_{r4}({\mathbf p^4\!+\! p^\prime\,^4}) \over {\mathbf k^2}}
    \!+\! \ldots ,\nn
\end{eqnarray}
\vskip-0.2truecm\noindent
where the total spin ${\mathbf S} = { (\bsigma_1 + \bsigma_2) / 2}$. $V^{(3)}$
could have a term of the form ${\mathbf p}^2{\mathbf p^\prime}^2/ {\mathbf
k^2}$, but it is more convenient to rewrite this as $({\mathbf p}^4 +{\mathbf
p^\prime}^4)/(2{\mathbf k^2})$ plus terms that vanish on-shell. In
Eq.~(\ref{Vpp}) all momentum dependence is explicit, and only the vector
coefficient ${\bf U}_\Lambda$ depends on spin. The odd terms, $V^{(2n+1)}$,
have coefficients of order $\alpha$ from tree level matching, while the even
terms, $V^{(2k)}$, are first generated at one-loop. Matching on-shell at
$\nu=1$:
\vskip-0.1truecm
\begin{eqnarray} \label{Ueo}
&& U_c(1) = -4 \pi Z \alpha \,,\quad
   U_k(1) = \frac{\pi^2 Z^2 \alpha^2\, m_R }{m_1 m_2} \,, \\
&& U_2(1) = \frac{\pi Z \alpha}{2} \Big(\frac{1}{m_1}-\frac{1}{m_2}\Big)^2
     \,,\ \quad
   U_s(1) = \frac{4\pi Z \alpha}{3 m_1 m_2} \,,\nn\\
&& {\mathbf U}_\Lambda(1) = \pi Z \alpha \Big[ \frac{\bsigma_1}{m_1^2}
     \!+\! \frac{\bsigma_2}{m_2^2} \!+\! \frac{4{\bf S}}{m_1 m_2} \Big], \quad
   U_r(1) = -\frac{4 \pi Z \alpha}{m_1 m_2}  . \nn
\end{eqnarray}
\vskip-0.1truecm\noindent
The reduced mass $m_R=m_1 m_2/(m_1+m_2)$. Terms not needed have been
omitted. Let $U_{2+s}\equiv U_2 + U_s {\mathbf S}^2$. Positronium has
additional contributions from one-photon annihilation, $U_{2+s}^\gamma(1)=\pi
\alpha {\mathbf S}^2/m_e^2$, and from two-photon and three-photon annihilation
graphs which give the imaginary terms (where $m_e$ is the electron mass):
\vskip-0.2truecm
\begin{eqnarray}
  U_{2+s}^{\gamma\gamma}=\frac{i\pi\alpha^2( {\mathbf S}^2\!-\!2)}{m_e^2} ,\quad
  U_{2+s}^{\gamma\gamma\gamma}=-\frac{i\pi \alpha^3 4(\pi^2\!-\!9)
   {\mathbf S}^2}{9\pi m_e^2} .
\end{eqnarray}
\vskip-0.15truecm

Evaluating the matrix element of a potential of order $\alpha^r v^s$ gives a
contribution to the energy at order $\alpha^{r+s+2}$. The iteration of two
potentials of order $\alpha^{r_1} v^{s_1}$ and $\alpha^{r_2} v^{s_2}$  gives a
potential of order $\alpha^{r_1+r_2} v^{s_1+s_2}$. Thus, $V^{(-1)}$, $V^{(0)}$,
$V^{(1)}$, $V^{(2)}$, and $V^{(3)}$ contribute starting at order $\alpha^2$,
$\alpha^4$, $\alpha^4$, $\alpha^6$, and $\alpha^6$ respectively.  The products
$V^{(0)}V^{(0)}$, $V^{(0)}V^{(1)}$, and $V^{(1)}V^{(1)}$ contribute starting at
order $\alpha^6$. In evaluating matrix elements $\alpha\sim v$, $V^{(-1)}\sim
1$, so the Coulomb potential cannot be treated as a perturbation.

The LO anomalous dimension for a potential coefficient of order $\alpha^r$ will
be defined by terms of order $\alpha^{r+1}$, the NLO anomalous dimension by
terms of order $\alpha^{r+2}$, etc., rather than by the conventional definition
in terms of the number of loops. In bound states where the typical momentum is
smaller than the electron mass, the Coulomb potential has no anomalous
dimension. The first terms that have an anomalous dimension are $V^{(0)}$ and
$V^{(1)}$. The LO anomalous dimension for $V^{(0)}$ and $V^{(1)}$ generate the
LO series in the energy $\alpha^4 (\alpha \ln \alpha)^n$. Their NLO anomalous
dimensions generate the NLO series $\alpha^5 (\alpha \ln \alpha)^n$. Matrix
elements of $V^{(2)}$ or $V^{(3)}$ or products of $V^{(0)}$ and $V^{(1)}$ first
contribute at order $\alpha^6 (\alpha \ln \alpha)^n$, the same order as the
NNLO anomalous dimensions for $V^{(0)}$ and $V^{(1)}$, and are not needed for
our analysis.  However, the potential $V^{(2)}$ mixes into $V^{(1)}$ at NLO,
and is therefore necessary for solving the VRG equations.

The anomalous dimensions in the effective theory can be generated by soft,
potential and ultrasoft loops, with energy-momentum of order $(k^0\sim
mv,{\mathbf k}\sim mv)$, $(k^0\sim mv^2,{\mathbf k}\sim mv)$ and $(k^0\sim
mv^2,{\mathbf k}\sim mv^2)$ respectively. There are no potential anomalous
dimensions at LO. In NRQED soft vertices are labelled by $\sigma \ge 1$, and a
loop graph with two soft vertices is of order $v^{ \sigma_1 + \sigma_2 -
1}$~\cite{LMR}. The one-loop soft anomalous dimension is non-zero only for
$V^{(2n-1)}$ with $n\ge 1$. At LO only the running of $V^{(1)}$ is needed 
(from Fig.~\ref{fig_LO}a with two $\sigma=1$ vertices):
\begin{figure}
 \centerline{\hbox{\epsfxsize=2cm \epsfbox{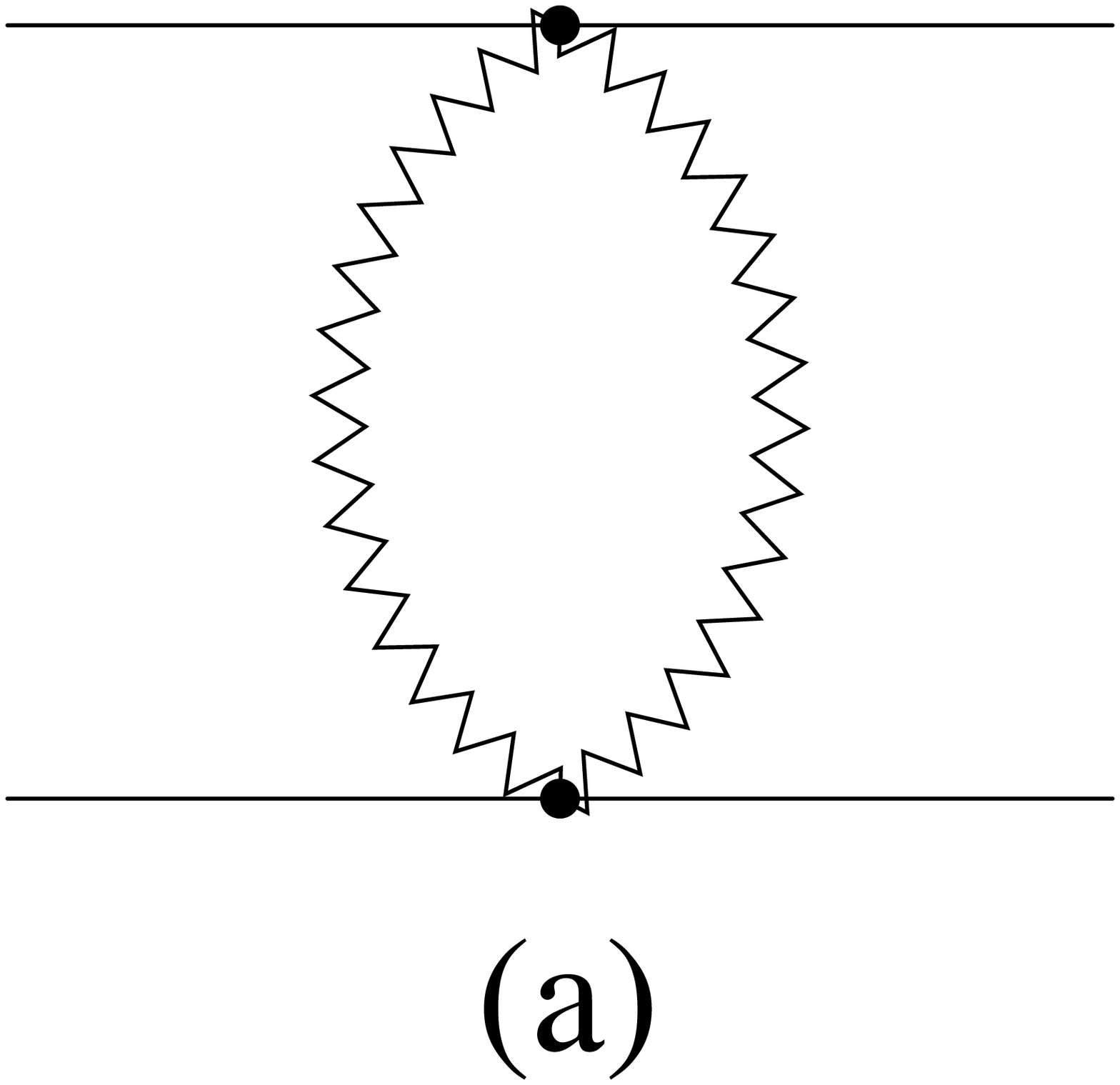}\qquad
 \epsfxsize=2cm\epsfbox{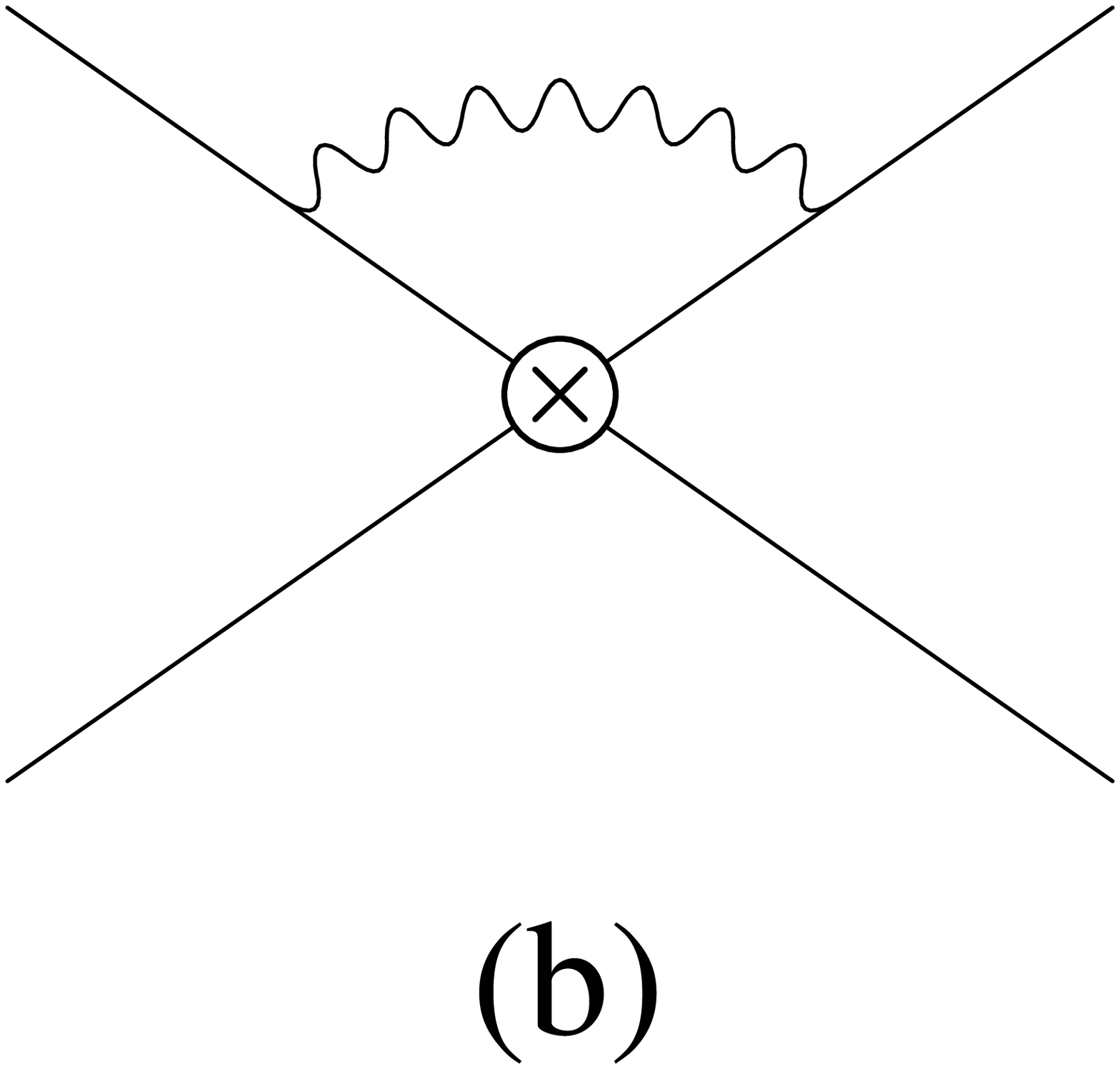}\qquad
 \epsfxsize=2cm\epsfbox{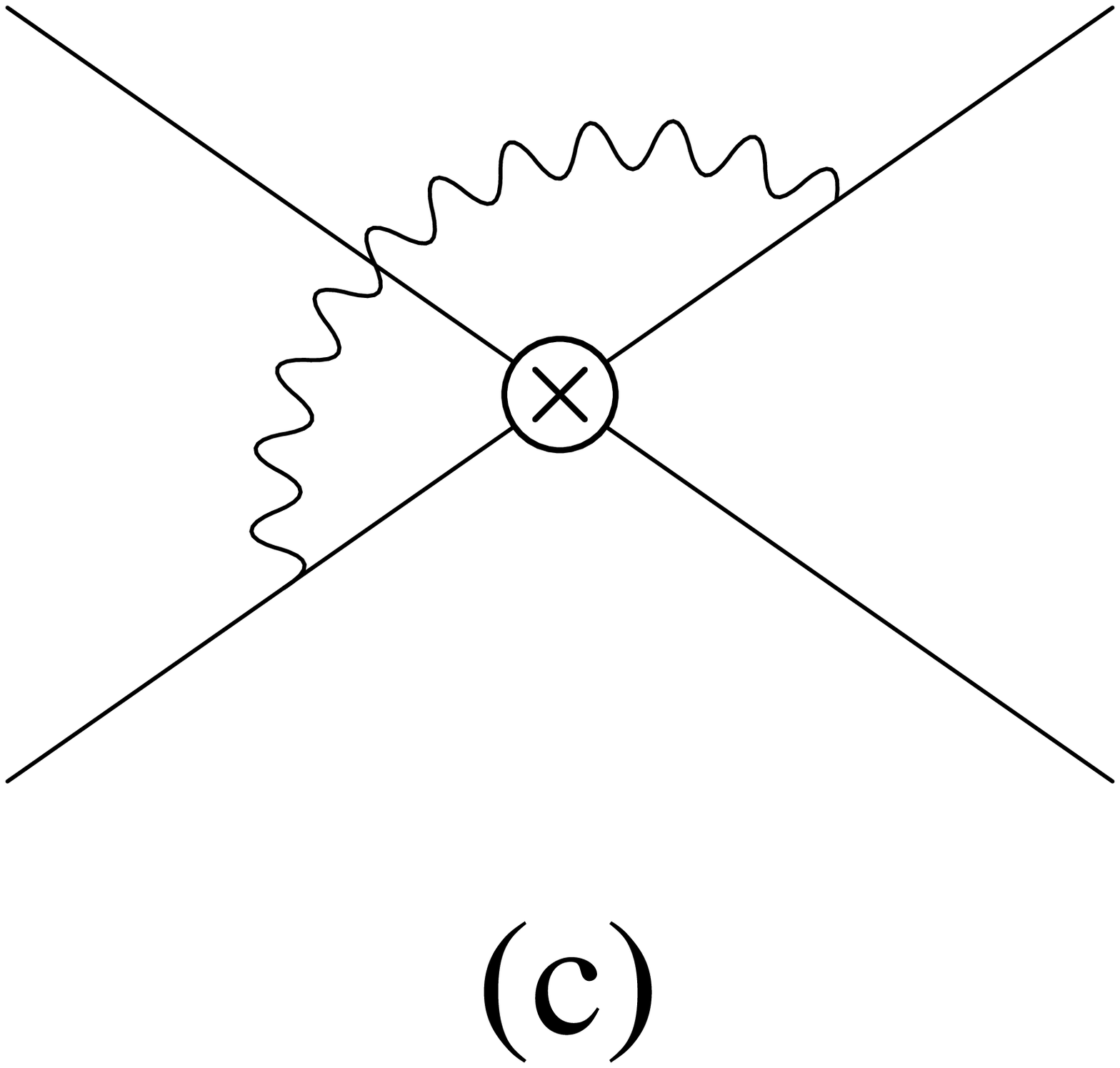}\raise1.4cm\hbox{$+\ldots$}}}
 {\tighten \caption{Graphs contributing to the soft (a) and ultrasoft (b,c,...)
 anomalous dimensions at leading order. The potential is denoted by $\otimes$, and one
 sums over all possible ultrasoft exchanges (including wavefunction
 renormalization).} \label{fig_LO} }
\end{figure}
\vskip-0.2truecm
\begin{eqnarray}\label{soft}
 &&   \nu {d U_2 \over d \nu} = {14 Z^2 \alpha^2 \over 3 m_1 m_2}  \,,\qquad
 \end{eqnarray}
where the other coefficients in $V^{(1)}$ have zero soft anomalous dimensions.
The LO ultrasoft anomalous dimension is independent of the momentum structure
of the potential (from Fig.~\ref{fig_LO}b, etc.):
\begin{eqnarray}\label{usoft}
  \nu {d  \over d \nu} V({\bf p,p'}) = {2\alpha \over 3 \pi}\left({1\over m_1}
  + {Z \over  m_2}\right)^2 \,  {\mathbf k}^2\: V({\bf p,p'})  \,.
\end{eqnarray}
At LO, the VRG equations for $V^{(0,1,2)}$ are:
\begin{eqnarray} \label{LOarray}
 \nu {d U_k \over d \nu} &=& 0 \,, \nn\\
 \nu {d U_2 \over d \nu} &=& {2\alpha \over 3 \pi}\left({1\over m_1}
   + {Z \over  m_2}\right)^2 U_c + {14 Z^2 \alpha^2 \over 3 m_1 m_2} \,,\nn\\
 \nu {d U_3 \over d \nu} &=& {2\alpha \over 3 \pi}\left({1\over m_1}
   + {Z \over  m_2}\right)^2 U_k + \gamma_1 U_c + \gamma_2 U_c^2 \,,
\end{eqnarray}
obtained by summing Eqs.~(\ref{soft}) and (\ref{usoft}). The last equation also
has additional spin-independent contributions denoted by $\gamma_{1,2}$, which
are not needed here. The $U_2$ integration is trivial,
\begin{eqnarray}\label{U2int}
 U_2\left( \nu \right) &=&  \gamma_0 U_c \ln \nu + U_2\left(1 \right) \,,\nn\\
  \gamma_{0} &=& {2 \alpha \over 3 \pi }\left( {1\over m_1^2} + {Z \over 4
    m_1 m_2} + {Z^2 \over m_2^2} \right),
\end{eqnarray}
and $U_2(1)$ has no large logarithm. The matrix element of the $U_2$ potential
is $(m_R Z \alpha)^3/\pi n^3$ for the $nS$ state (and zero for $L\ne 0$
states). With the coefficient $U_2(\nu=Z\alpha)$ from Eq.~(\ref{U2int}) this
gives the logarithmic energy shift
\begin{eqnarray}\label{bethe}
\Delta E = -{8 Z^4 \alpha^5 m_R^3 \over 3 \pi n^3}\left( {1\over m_1^2} +
 {Z \over 4 m_1 m_2} + {Z^2 \over m_2^2} \right)  \ln Z \alpha , \nn
\end{eqnarray}
which is the well-known Bethe logarithm in the Lamb shift, and is valid for
Hydrogen, muonium and positronium. Equation~(\ref{U2int}) has no imaginary part
or spin-dependence, so there is no contribution to the decay width or hyperfine
splitting at this order. Equation~(\ref{bethe}) has been computed before using
an effective field theory~\cite{PSLamb}. The VRG method makes it clear that the
LO series $\alpha^4 (\alpha \ln \alpha)^n$ has only a single term---the
anomalous dimension for $U_2$ depends on $\alpha$ and $U_c$, both of which do
not run, so integrating the VRG equation produces only a $\ln \nu$ term. Below
the electron mass, QED is not very efficient at generating logarithms.

At NLO, one needs the anomalous dimension of $V^{(0)}$ to order $\alpha^4$, and
of $V^{(1)}$ to order $\alpha^3$. The possible terms that can contribute are
determined by using the $v$-counting formula in Ref.~\cite{LMR}. The NLO
anomalous dimension of $V^{(0)}$ has the form $\nu dU_k/d\nu=\gamma_3 U_c$,
where $\gamma_3$ is generated by Fig.~(\ref{fig_NLO})b iterated with $U_c$. The
NLO anomalous dimension for $V^{(1)}$ has several contributions from potential
loops (see Fig.~\ref{fig_NLO}a): $U_c^3$ with a $v^4$ insertion, $U_c^2
V^{(1)}$ with a $v^2$ insertion, $U_c V^{(1)}V^{(1)}$, $U_c^2 V^{(3)}$, $U_c
V^{(0)}$ with a $v^2$ insertion, $U_c V^{(2)}$, and $V^{(1)}V^{(0)}$. There is
no ultrasoft anomalous dimension for $V^{(1)}$ at NLO, but there is a soft
contribution shown in Fig.~\ref{fig_NLO}b with two $\sigma=1$ interactions,
which has the form $\rho_s Z^3 \alpha^3/m_1 m_2$. $\gamma_{3}$ and $\rho_s$
do not contribute to our results. The only NLO anomalous dimension needed is:
\begin{figure}
\centerline{\epsfxsize=2.8cm\epsfbox{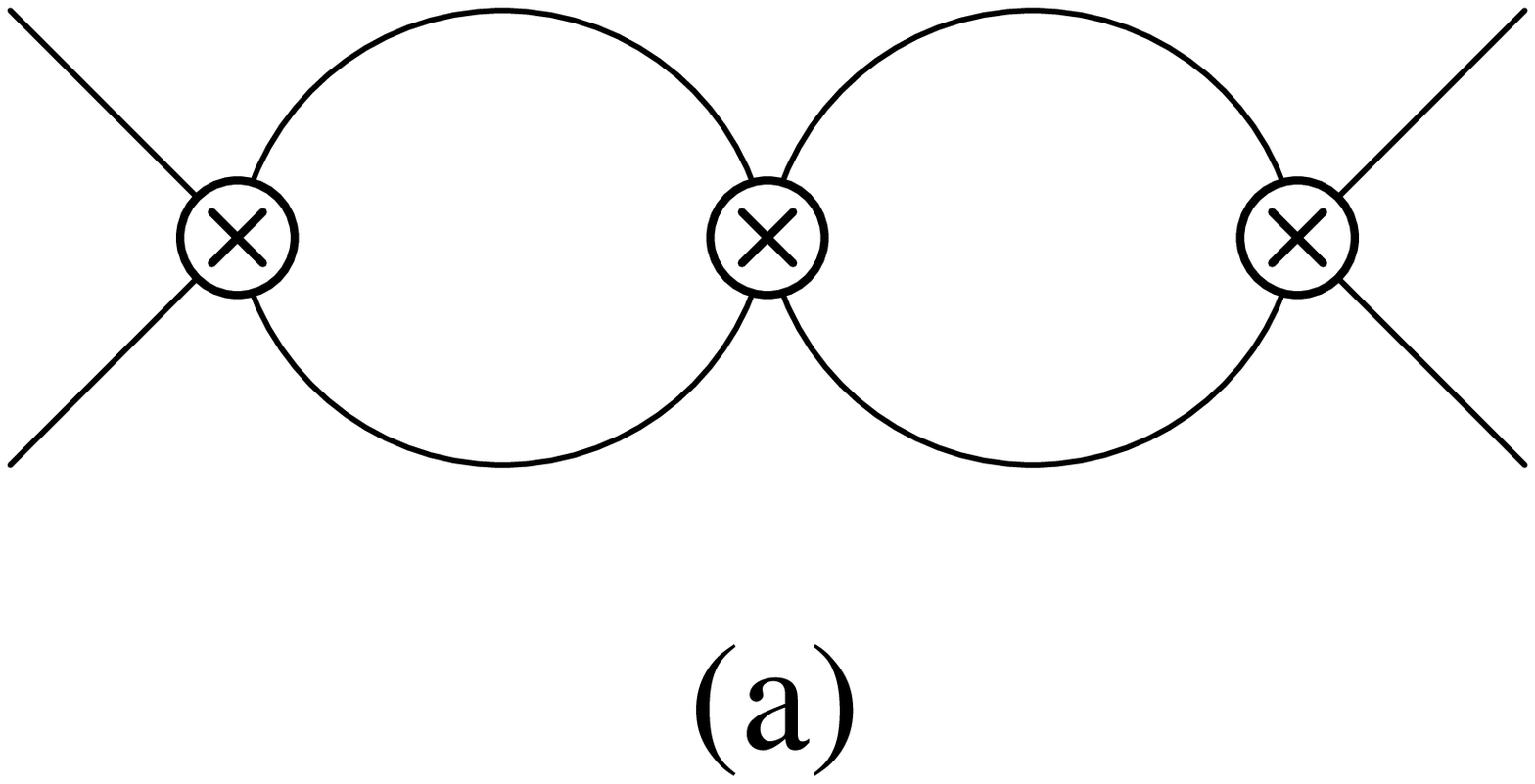} \qquad\quad
  \epsfxsize=2cm\epsfbox{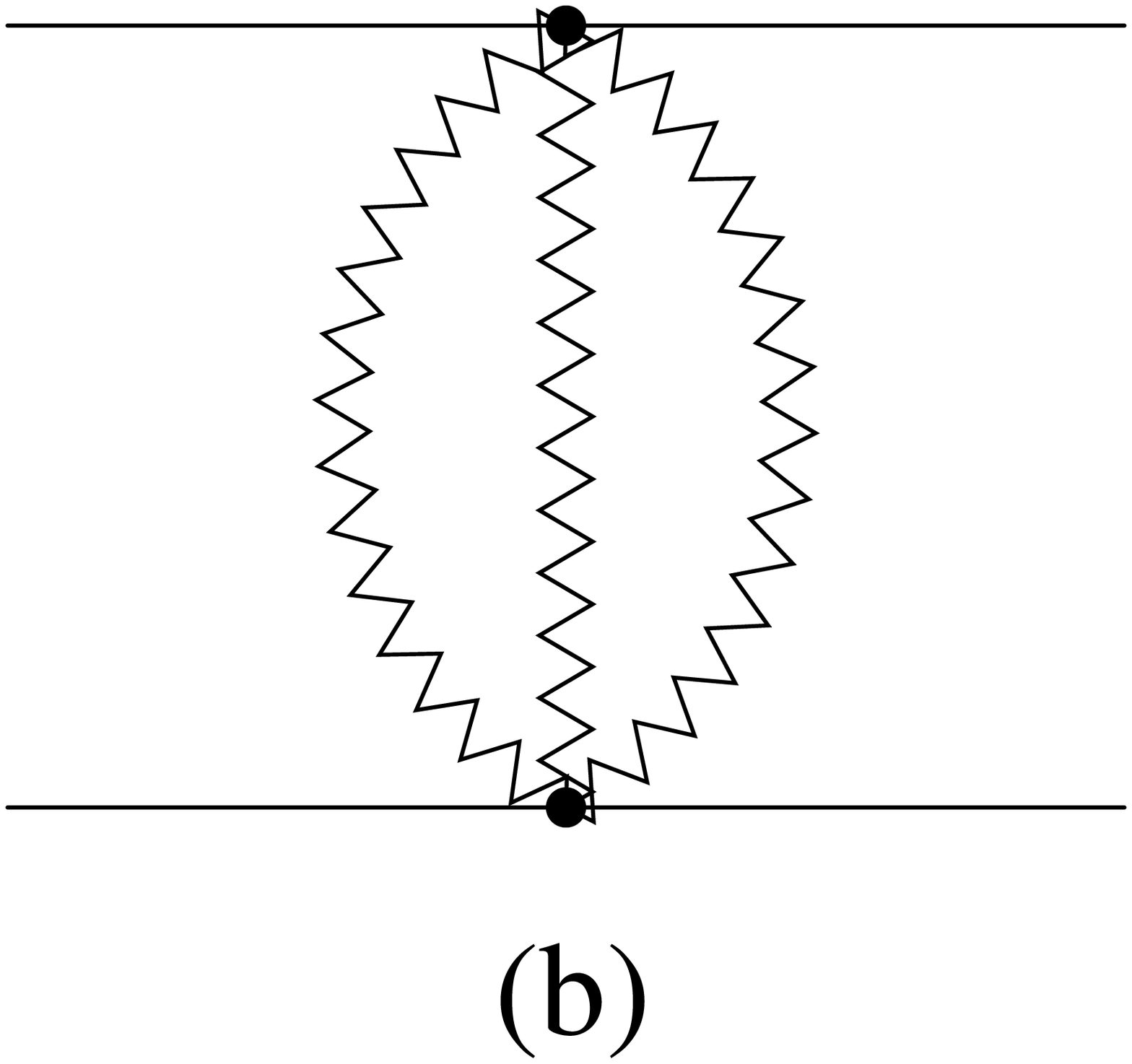}  }
 {\tighten \caption{Examples of graphs contributing to the NLO anomalous
 dimensions. The $\otimes$ denotes insertions of terms in the potential such
 as $U_c$, $U_2$, etc.} \label{fig_NLO} }
\end{figure}
\begin{eqnarray}\label{NLO}
 &&  \left. \nu {d U_{2+s} \over d\nu}  \right|_{\rm NLO} =
   \rho_{ccc}\, U_c^3 + \rho_{cc2}\, U_c^2  \left( U_{2+s}+ U_r
   \right) \nn\\
 &&\quad + \rho_{c22}\, U_c\left(U_{2+s}^2 +2 U_{2+s} U_r + \frac34 U_r^2
   -9 U_t^2 {\bf S^2} \right)\nn\\
 && \quad +\rho_{ck}\, U_c U_k +\rho_{k2}\, U_k \left(U_{2+s}  +
   U_r/2\right) \nn\\
 && \quad + \rho_{c3}\, U_c \left(U_3+\ldots\right)
   + \rho_s {Z^3 \alpha^3 \over m_1 m_2},
\end{eqnarray}
where $U_{2+s}=U_2 + U_s {\mathbf S}^2$ and
\begin{eqnarray}
 \rho_{ccc}&=& -{m_R^4 \over 64 \pi^2} \left( {1\over m_1^3} + {1\over
   m_2^3}\right)^{\!2},\qquad
 \rho_{c22}= -{m_R^2 \over 4 \pi^2},\nn\\
 \rho_{cc2}&=& -{m_R^3 \over 8 \pi^2} \left( {1\over m_1^3} +
   {1\over m_2^3} \right),\qquad
 \rho_{c3} = { 2 m_R \over  \pi^2},\nn\\
 \rho_{ck} &=& {m_R^2 \over 2 \pi^2}\left({1\over m_1^3} +{1\over m_2^3}
    \right), \qquad
 \rho_{k2} = { 2 m_R \over  \pi^2}.
\label{NLOvalues}
\end{eqnarray}
The potential $V^{(3)}$ does not mix into $V^{(1)}$ when the basis is chosen to
avoid the ${\mathbf p}^2{\mathbf p^\prime}^2/{\mathbf k}^2$ term.
Equation~(\ref{NLO}) depends on $V^{(2)}$, so we also require its LO VRG 
equation in Eq.~(\ref{LOarray}). The ellipses in Eq.~(\ref{NLO}) denote terms
other than ${U}_3$ in the $V^{(2)}$ potential (e.g. $U_{3s}$), which contribute
to the running of $U_2$.  However, they do not have a LO anomalous dimension,
and thus contribute with one less logarithm and are not required for our
analysis. 

Integrating Eq.~(\ref{NLO}) generates an infinite series of terms, since it is
non-linear in $U_{2+s}$, including $\alpha^2 (\alpha\ln \nu)^n$ terms with
$n=1,2,3$. Thus, the NLO $\alpha^2 (\alpha\ln \nu)^n$ series terminates
after three terms. Integrating the $U_c U_2^2$ term in Eq.~(\ref{NLO}) gives
these three possible terms, depending on whether each $U_2$ is replaced by the
$\gamma_0 U_c \ln \nu$ or $U_2(1)$ term in Eq.~(\ref{U2int}).

\begin{figure}
\centerline{ \hbox{\epsfxsize=3cm\epsfbox{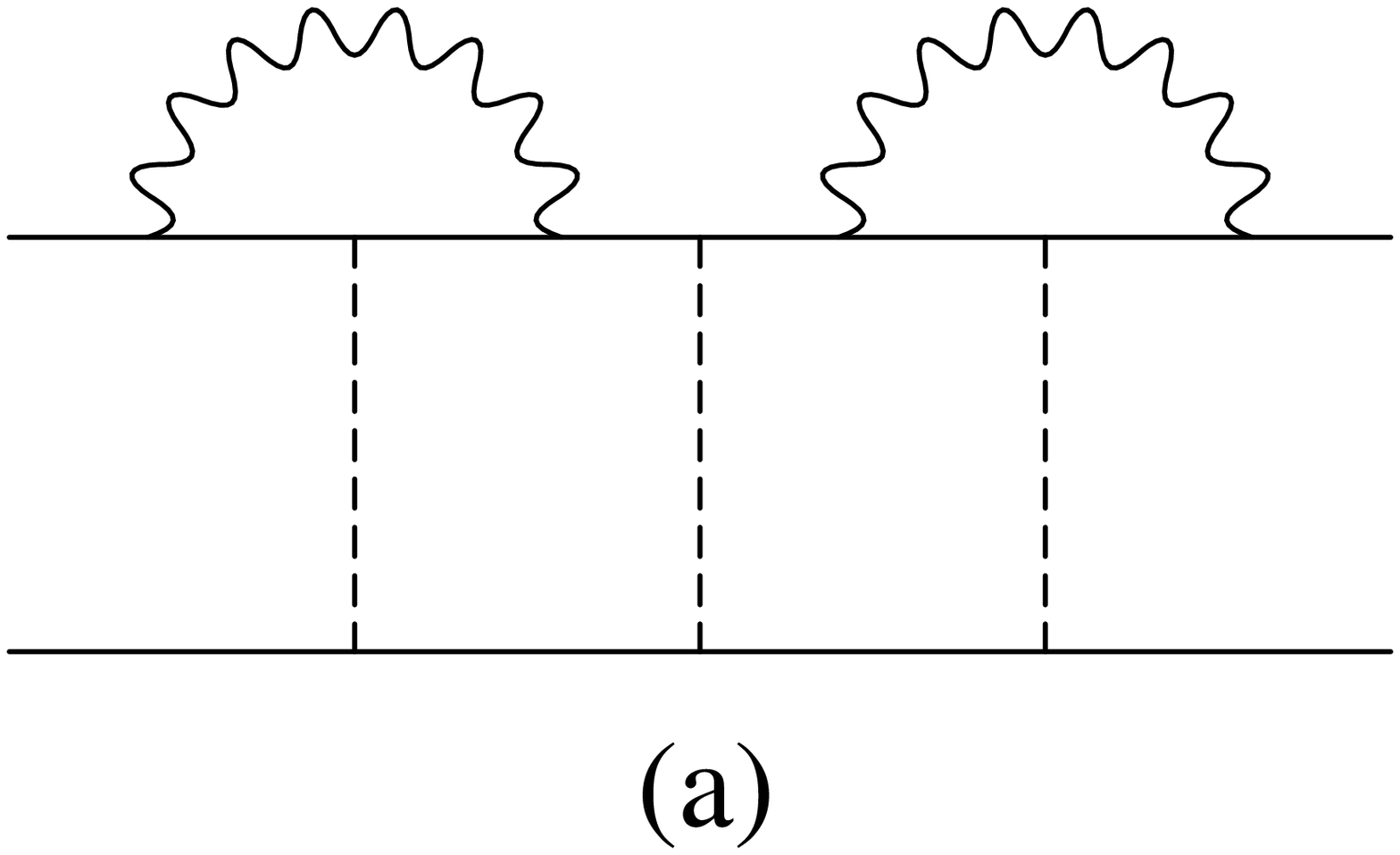}\qquad
 \epsfxsize=3cm\epsfbox{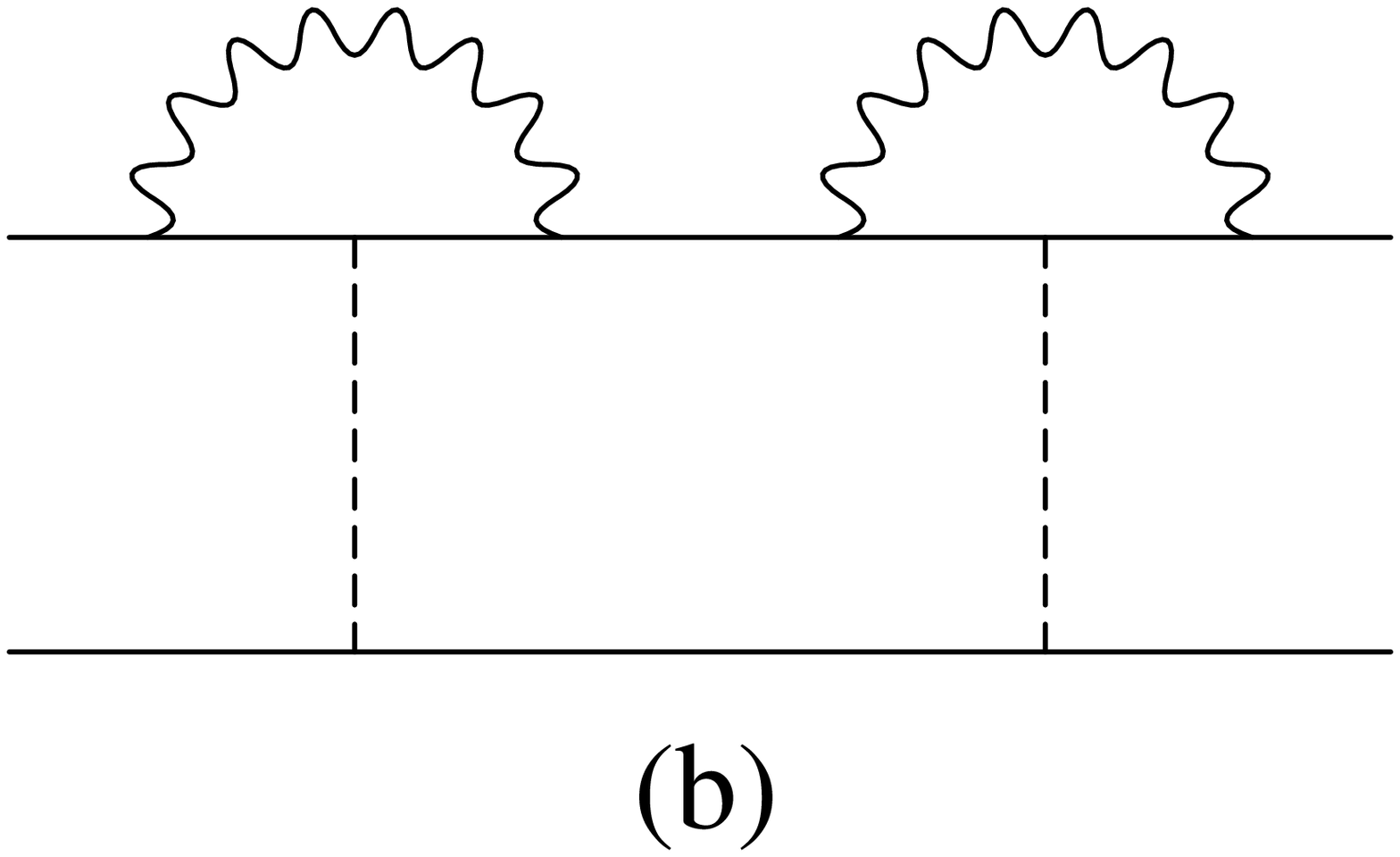}}} \smallskip
{\tighten \caption{Fig.~(a) is the $\alpha^8 \ln^3\! \alpha$ contribution to the
Lamb shift computed by Karshenboim. Fig.~(b) is the additional contribution of
Yerokhin. The dashed lines are Coulomb exchanges.} \label{fig_yerokhin} }
\end{figure}

The simplest term is the $\ln^3\! \nu$ term in $U_{2+s}(\nu)$:
\begin{eqnarray}\label{11}
{1\over 3}\, \gamma_0^2\, \rho_{c22}\, {U}_c^3(1)\, \ln^3\nu,
\end{eqnarray}
whose tree-level matrix element determines the $\alpha^8 \ln^3\! \alpha$ Lamb
shift for the $nS$ state
\begin{eqnarray}
\Delta E = {64 m_R^5 \alpha^8 Z^6 \over 27 \pi^2 n^3}\left(
{1\over m_1^2} + {Z\over 4 m_1 m_2} + {Z^2\over  m_2^2} \right)^2
\ln^3\! \left(Z \alpha \right).\nn
\end{eqnarray}
The $\alpha^8 \ln^3\! \alpha$ term is spin-independent and has no imaginary
part, so it does not contribute to the hyperfine splitting or the positronium
lifetime. The structure $\gamma_0^2 p_{c22}$ in Eq.~(\ref{11}) is the same as
the matrix element in~Fig.~\ref{fig_yerokhin}a  computed by
Karshenboim~\cite{karshenboim}, and our results reduces to his for Hydrogen
($m_2\to\infty$). For positronium, we find
$\Delta E = {3 m_e \alpha^8 \ln^3\!  \alpha/ (8 \pi^2 n^3)}$.
In the effective theory, the matrix element (for the log terms) becomes the
products of the graphs in Fig.~\ref{fig_NLO}a and Fig.~\ref{fig_LO}.
Yerokhin~\cite{yerokhin} has an additional contribution
Fig.~\ref{fig_yerokhin}b, which has the structure $\gamma_0^2$ times the
${U}_2^2$ matrix element. While the loop graph with two ${U}_2$'s is linearly
divergent, it is finite in dimensional regularization and does not contribute
to the anomalous dimension or to the $\alpha^8 \ln^3\! \alpha$ term. 

The piece of the $\ln^2 \nu$ term that depends on the matching
value for ${U}_{2+s}$ at $\nu=1$ is
\begin{eqnarray}\label{NLOhfs}
 \gamma_0\, \rho_{c22}\, {U}_c^2(1) \left[{U}_2(1)+ {U}_s(1) {\mathbf S}^2
  \right] \ln^2 \nu.
\end{eqnarray}
The remaining terms depend on the matching values for $U_c$ and $U_k$, are
spin-independent, and only contribute to the Lamb shift.  The matching value is
${U}_{s}(1)=4  \pi Z \mu_1 \mu_2 \alpha /(3 m_1 m_2 )$ where $\mu_i$ are the
magnetic moments in units of $e/(2m_i)$. Using Eq.~(\ref{NLOhfs}) and taking
the difference between the energies for ${\mathbf S}^2=2$ and ${\mathbf S}^2=0$
gives the $\alpha^7 \ln^2 \alpha$ hyperfine splitting for muonium and Hydrogen
for the $nS$ state:
\begin{eqnarray}
  \Delta E = -{64 Z^6 \alpha^7 m_R^5 \mu_1 \mu_2 \over 9m_1m_2 \pi n^3} 
  \left[{1\over m_1^2} \!+\! {Z\over 4 m_1 m_2} \!+\! {Z^2\over  m_2^2} 
  \right]\ln^2 (Z \alpha),\nn
\end{eqnarray}
in agreement with Refs.~\cite{karshenboim,kinoshita}. The matching value for
positronium, $U_s(1) = 7\pi \alpha /(3 m_e^2)$, and $\mu_1=\mu_2=1$ gives the
hyperfine splitting for the $nS$ state:
\begin{eqnarray}
  \Delta E = -{7  m_e \over 8 \pi n^3}\: \alpha^7 \ln^2\! \alpha ,
\end{eqnarray}
in agreement with Refs.~\cite{karshenboim,thesis,melnikov}. 

The positronium width is
$\Gamma=-2\hbox{Im}\, E$. The matching coefficients ${U}_{2+s}$ have imaginary
parts from the two- and three-photon annihilation graphs, of order $\alpha^2$
and $\alpha^3$ respectively. They are the lowest order contributions to the
imaginary part,  so we can use Eq.~(\ref{NLOhfs}) for the widths, even
though real parts of the same order have been neglected:
\begin{eqnarray}
 {\Delta \Gamma\over \Gamma_0} = \gamma_0\, \rho_{c22}\, {U}_c(1)^2 \ln^2 \nu =
  -{3  \over 2 \pi}\alpha^3 \ln^2\! \alpha,
\end{eqnarray}
for the ortho- and para-positronium widths in agreement with
Ref.~\cite{karshenboim}. The $\alpha^7 \ln^2\! \alpha$ Lamb shift depends on
$\gamma_{1,2}$, and will be discussed elsewhere.

The $\ln \nu$ term that depends on the matching value of $U_{2+s}$ is
\begin{eqnarray}
&& U_{2+s} \Big[ \rho_{c22} U_c \left( {U}_{2+s} +2  U_r \right) + \rho_{cc2}
U_c^2 + \rho_{2k} U_k \Big] \ln \nu , \nn
\end{eqnarray}
where the $U_i$'s are evaluated at $\nu=1$. 
Using Eqs.~(\ref{NLOvalues}) and (\ref{Ueo}), the imaginary part gives the
positronium width:
\begin{eqnarray}
{\Delta \Gamma \over \Gamma_0}= \left({m_e^2 \over 2 \pi} \hbox{Re}\, U_{2+s}
 - 2 \right)\ln \nu  = \left( {7 {\mathbf S}^2 \over 6} - 2 \right)
 {\alpha^2} \ln \alpha, \nn
\end{eqnarray}
so for ${\mathbf S}^2=2$ and ${\mathbf S}^2=0$ we have:
\begin{eqnarray}
\left({\Delta \Gamma \over \Gamma_0}\right)_{\rm ortho} =
  {\alpha^2 \over 3}\ln \alpha \,, \qquad
\left({\Delta \Gamma \over \Gamma_0}\right)_{\rm para} =
  -2{\alpha^2}\ln \alpha \,, \nn
\end{eqnarray}
in agreement with Ref.~\cite{caswell2}. The $\alpha^6 \ln \alpha$ Lamb shift 
and hyperfine splitting have contributions from $V^{(2)}(\nu=1)$, $\gamma_3$ 
and $\rho_s$.

There are two infinite series of logarithmic terms that are easily identified.
Neither of them gives the complete contribution at a given order, but they do
show that there are logarithmic terms of arbitrarily high order.
The ultrasoft anomalous dimension Eq.~(\ref{usoft}) generates the potential
\begin{eqnarray}
  V(\nu=Z \alpha) = \exp\left[ {2 \alpha \over 3 \pi}\left({1\over m_1} +
  {Z \over m_2}\right)^2 {\mathbf k}^2 \ln Z \alpha \right] {U_c(1) \over
  {\mathbf k}^2},\nn
\end{eqnarray}
where the exponential is related to the Sudakov form factor.  In position space,
the modified Coulomb potential is
\begin{eqnarray}\label{esudakov}
  V(Z\alpha)=-{Z \alpha \over r} {\rm Erf}\left[r \sqrt{3 \pi \over 8 \alpha
  \ln[1/(Z\alpha)]} \left({1\over m_1}\!+\! {Z \over m_2}\right)^{-1} \right],\nn
\end{eqnarray}
where ${\rm Erf}$ is the error function. This gives a series in the energy of
the form $\alpha^5 \ln Z \alpha (\alpha^3 \ln Z\alpha)^{n/2}$, $n\ge0$. 
The second infinite
series of logarithmic terms is obtained by integrating the VRG equation for
$U_2$ retaining only the $U_2$ terms in Eqs.~(\ref{LOarray}) and (\ref{NLO}),
\begin{eqnarray}
  \nu {d U_2 \over d \nu}= \gamma_0 U_c + \rho_{c22} U_c U_2^2 .
\end{eqnarray}
Solving this equation gives
\begin{eqnarray}
 U_2(\nu)\!=\! {U_2(1)\!+\! \sqrt{\gamma_0/|\rho_{c22}|}\tanh\left[\sqrt{\gamma_0
  |\rho_{c22}}| \, U_c(1) \ln \nu\right]\over
  1\!+\!\sqrt{ |\rho_{c22}| /\gamma_0} U_2(1) \tanh\left[
  \sqrt{\gamma_0 |\rho_{c22}|}\,  U_c(1) \ln \nu\right]}  , \nn
\end{eqnarray}
which has an expansion of the form $\alpha^2 \ln \alpha (\alpha^3 \ln^2
\alpha)^n$. The first two terms are the $\alpha^5 \ln \alpha$ and $\alpha^8
\ln^3\! \alpha$ Lamb shifts. The difference between the exact expression and the
first two terms is less than 1~Hz in the Lamb shift.

We have derived logarithmic terms in QED bound states using the VRG. The
results are in agreement with results computed previously using other methods.
The computation of the $\alpha^8 \ln^3\! \alpha$ Hydrogen Lamb shift by the VRG
supports the value computed by Karshenboim~\cite{karshenboim}. We have also
computed the $\alpha^8 \ln^3\! \alpha$ positronium Lamb shift. The VRG approach
makes the universal nature of the $\ln \alpha$ terms clear, and is also an
efficient way of computing these terms. The computations in this paper support
a key feature of the VRG method proposed in Ref.~\cite{LMR}: the simultaneous
running of soft and ultrasoft effects from $m$ using a subtraction velocity. We
have also identified two infinite series of logarithms in QED.

We thank A.~Hoang and K.~Melnikov for discussions.  This work was supported in
part by the DOE grant DOE-FG03-97ER40546, and by NSF award PHY-9457911.
\vskip-0.2truecm
{\tighten

} %end tighten (references & figure captions)

\end{document}